\documentclass[aps,prl,twocolumn]{revtex4-1}

\usepackage{graphicx}
\usepackage{latexsym}
\usepackage{amsmath}
\usepackage{amssymb}
\usepackage{amsfonts}
\usepackage{color}

\begin{document}



\title[Stable ``antiferromagnetic'' vortex lattice]{Stable ``antiferromagnetic'' vortex lattice imprinted into a type-II superconductor}

\author{V N Gladilin$^{1,2}$, J Tempere$^{2}$, J T Devreese$^{2}$ and V V Moshchalkov$^{1}$}
\address{$^1$ INPAC -- Institute for Nanoscale Physics and Chemistry,
Katholieke Universiteit Leuven, Celestijnenlaan 200D, B--3001
Leuven, Belgium}
\address{$^2$ TQC -- Theory of Quantum and Complex
Systems, Universiteit Antwerpen, Universiteitsplein 1, B-2610
Antwerpen, Belgium}

\begin{abstract}
In type-II superconductors, where vortices and antivortices tend to
annihilate, only a ``ferromagnetic'' vortex lattice, with the same
orientation of vortex magnetic moments, is usually formed in a
homogeneous external magnetic field. Using the time-dependent
Ginzburg-Landau formalism, we demonstrate that a checkerboard
vortex-antivortex lattice (``antiferromagnetic vortex lattice''),
imprinted onto a superconducting film by a periodic array of
underlying clockwise and counterclockwise microcoils generating
spatially periodic positive and negative magnetic field pulses and
then trapped by an array of artificial pinning centers, remains
stable even after the imprinting magnetic field pulse is switched
off.
\end{abstract}


\maketitle

Magnetic moments in solids can order ferro- or
antiferromagnetically~\cite{Weiss,Heisenberg,Neel}. In a certain
analogy with that, in type-II superconductors the ordered vortex
state can be built up from entities of the same or the opposite
polarity - vortices and antivortices. Such vortex lattices can be
referred to as ``ferromagnetic'' and ``antiferromagnetic'' vortex
states. In this terminology, the Abrikosov vortex
lattice~\cite{Abrikosov} is an example of a ``ferromagnetic'' state.
During the last decade, vortex patterns consisting of
superconducting vortices and antivortices have attracted significant
theoretical and experimental efforts. A number of stable
vortex-antivortex (VAV) configurations have been predicted and
experimentally detected in symmetric mesoscopic superconducting
samples subjected to a homogeneous magnetic
field~\cite{Chibotaru00,Misko03,Geurts06}. VAV pairs can be
naturally created and stabilized by spatially inhomogeneous fields
of micromagnets~\cite{Lyuksyutov98,lange03,Kayali04,Erdin05},
leading to various commensurability effects and novel stable VAV
configurations in superconductor-ferromagnet hybrids with regular
arrays of magnetic dipoles~\cite{lange03,priour04,milosevic06}.

Whereas VAV lattices were induced using an array of micromagnets,
the creation of a VAV state that remains stable without permanent
presence of external spatially-alternating magnetic field is still
an open problem. An appealing idea relying on superconducting
relaxation dynamics was formulated in
Refs.~\cite{Ghinovker98,Ghinovker99}. This proposal considers
vortices and antivortices that are spontaneously generated during
the recovery of superconductivity after a sample was heated locally
with a laser pulse. In the presence of strong pinning centers some
of those vortices and antivortices can be frozen and persist
practically indefinitely. In this context it was
suggested~\cite{Ghinovker01} that an analysis of the superconducting
phase recovery accompanied by the creation of vortices and
antivortices, provides a convenient tool to test the Kibble-Zurek
cosmological scenario~\cite{Kibble76,Zurek96} for nucleation of
topological defects during a symmetry-breaking phase transition.

In this Letter, we consider the use of a specially designed pulse of
spatially periodic magnetic field for imprinting a regular VAV
lattice, which, as we predict, remains stable after the removal of
the external magnetic field used for VAV imprinting. In particular,
we will demonstrate that due to mutual cancelation of forces, which
act on a vortex (antivortex) in the equilibrium state of such a
lattice, the antiferromagnetic vortex state can be stabilized even
in the case when pinning centers in the superconductor are
relatively weak.

To describe the vortex dynamics, we apply the time-dependent
Ginzburg-Landau (TDGL) formalism as presented in
Refs.~\cite{Chapman96,Silhanek11}. For the sake of convenience, we
use dimensionless variables by expressing lengths in units of
$\sqrt{2}\xi$ and currents in units of $\Phi_0\xi/(\sqrt{2}\pi \mu_0
\lambda^2)$, where $\Phi_0=\pi\hbar/e$ is the magnetic flux quantum,
$\mu_0$ is the vacuum permeability, $\xi$ is the coherence length
and $\lambda$ is the penetration depth. The unit of magnetic field,
$\Phi_0/(4\pi \xi^2)$, is half the second critical field. Our unit
of time is $2u\tau_{\rm GL}$, where $\tau_{\rm GL}$ is the
Ginzburg-Landau relaxation time and $u = \pi^4/[14\zeta(3)]$ with
$\zeta(x)$, the Riemann zeta function.

We consider a thin infinite superconductor film subjected to a pulse
of spatially periodic negative and positive magnetic fields
[Fig.1(a)]. In the model, this field is generated by an underlying
square array of circular quasi-one-dimensional current loops with
diameter $D$, located at a distance $d_1$ below the middle of the
superconductor layer. The current direction changes from one loop to
another in the checkerboard order as shown in Fig.~1(a), thus
generating a periodic array of local magnetic fields of the opposite
polarities. It is assumed that due to an appropriate arrangement of
the leads connecting the loops to each other, the magnetic field
induced by these leads is negligible. In the present calculations,
the spatial period of the structure is $L = 15$, the superconductor
film thickness is $d = 0.3$ and the Ginzburg-Landau parameter of the
superconductor is $\kappa\equiv \lambda/\xi= 1$. The current pulse
in the loops has trapezoidal shape: the current $I(t)$ linearly
increases from 0 to $I_{\rm max}$ in the time interval from $t = 0$
to $t = t_1$, remains constant $I(t) = I_{\rm max}$ in the time
interval $(t_1,\ t_2)$ and then linearly decreases down to 0 within
the interval $(t_2,\ t_3)$. A periodic lattice of up and down
out-of-plane magnetic fields, induced by such a current pulse,
creates in the superconducting layer a square lattice of vortices
and antivortices located just above the loops with counterclockwise
and clockwise current directions, respectively. Vortices and
antivortices, arranged to form a periodic checkerboard-like square
lattice, remain in equilibrium even after completely switching off
the applied magnetic field. This equilibrium is obviously unstable
in an ideally uniform superconducting film: an arbitrarily weak
perturbation would destroy the lattice due to the vortex-antivortex
attraction and subsequent recombination of the VAV pairs. However,
in reality superconducting films often contain numerous naturally
occurring pinning centers, which may prevent such a collapse of the
VAV lattice. Alternatively, in a more controlled way, a periodic
array of artificial pinning centers can be intentionally induced,
e.g., in the form of a periodic array of ``blind'' or complete
antidots~\cite{Baert95,Harada96}.
\begin{figure}
\centering
\includegraphics*[width=0.95\linewidth]{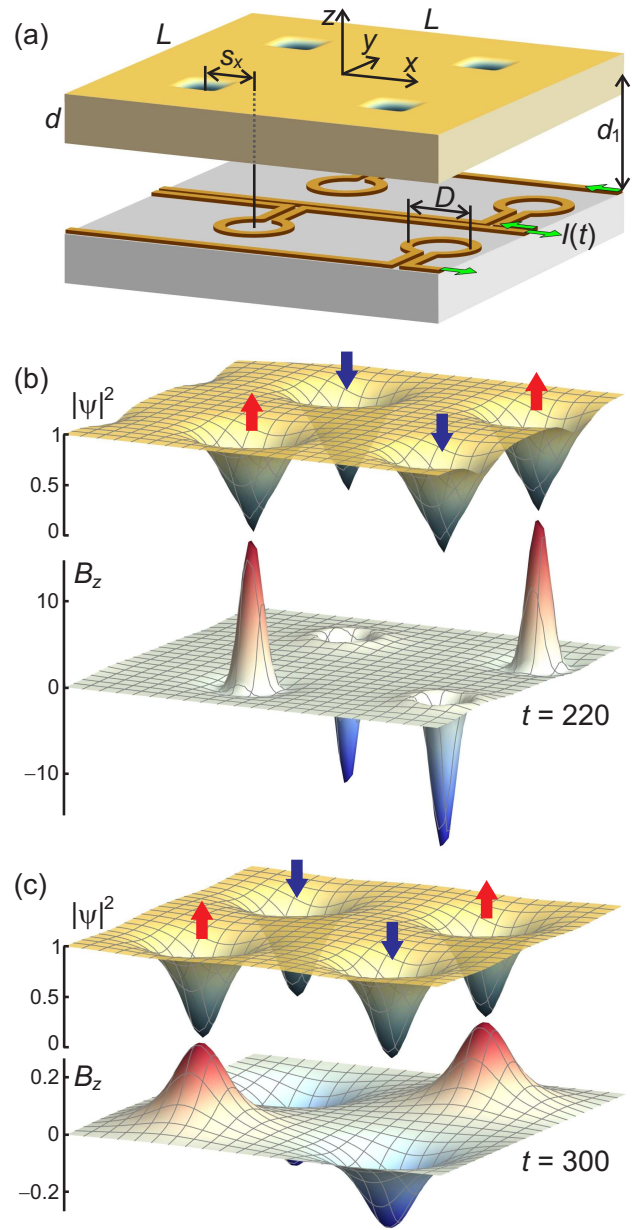}
\caption{(a) Unit cell of a periodic structure, which contains a
superconducting film and a square array of $\Omega$-shaped current
loops of diameter $D$ at a distance $d_1$ from the film. When
applying a current $I(t)$ as shown with green arrows, the magnetic
field induced by the loops alternates in checkerboard order. Spatial
modulations of the superconductor thickness provide an array of
vortex-pinning centers, which are shifted along the $x$-axis by a
distance $s_x$ with respect to the centers of the nearest current
loops. (b) At the end of the pulse plateau a lattice of vortices and
antivortices, seen as local suppressions in the square modulus of
the order parameter $|\psi|^2$, is created by the magnetic field
$B_z$ of the loops. The red (blue) arrows indicate the direction of
the magnetic moment of vortices (antivortices). (c) When the applied
current decreases and eventually vanishes, the vortices and
antivortices drift towards the nearest pinning centers. This results
in a stable VAV lattice characterized by an ``antiferromagnetic''
distribution of the induced magnetic field $B_z$. \label{Figure1}}
\end{figure}

Here we consider square array of artificial pinning centers formed
by indentations of depth $\delta\ll d$ made in the superconducting
film of thickness $d$ (blind antidots), as shown in Fig.~1(a). The
number and periodicity of the pinning sites corresponds to the
number and periodicity of the current loops, but the arrays of loops
and blind antidots are shifted with respect to each other. We have
analyzed shifts along the side of a unit cell, over a distance
denoted by $s_x$, and shifts along the diagonal of the unit cell,
over a distance $s_d$.

Figures~1(b) and 1(c) illustrate the formation of a VAV lattice in
the case of $\delta=0.06$, $d_1=0.5$, $D=1$, $s_x=2$, $I_{\rm
max}=4$, $t_1=10$, $t_2=210$, and $t_3=250$. Ramping up the
periodical array of up and down magnetic fields during the time
interval $(0,\ t_1)$ strongly suppresses the order parameter
$|\psi|$ just above the centers of the current loops. Due to the
radial gradient of the magnetic field, induced by a current loop, a
VAV pair is generated at each loop. This process is quite similar to
the VAV pair generation by a magnetic dipole considered in
Ref.~\cite{Gladilin09}. As the current pulse reaches its plateau,
some recombinations take place: the vortex induced by the field of a
current loop with counterclockwise current remains pinned to this
loop, while the corresponding antivortex is driven away from the
loop and recombines with a vortex pushed away from one of the
neighboring loops with clockwise current. After that, single
vortices and antivortices, which correspond to one flux quantum
$\Phi_0$ each, are left at the sites of the counterclockwise and
clockwise current loops, respectively. Due to the aforedescribed
processes, a VAV lattice, pinned to the current-loop array, is
imprinted into the superconductor [see Fig. 1(b)]. This lattice
pertains till the end $(t = t_2)$ of the plateau in the applied
field pulse. During the time interval $(t_2,\ t_3)$, when the
applied magnetic field decreases and eventually vanishes, the
vortices and antivortices gradually drift towards the nearest
pinning sites, resulting in the formation of an
``antiferromagnetic'' vortex state, which is shown in Fig.~1(c) and
which remains stable for an arbitrarily long time after the applied
field has been switched off.

Figure 2 summarizes the results of the TDGL simulations for the
structure shown in Fig.~1(a) at $\delta=0.06$, $d_1=0.5$, $D=1$, ,
$t_1=10$, $t_2=210$, and $t_3=250$. Depending on the magnitude of
the current in the loops, $I_{\rm max}$, and the horizontal ($s_x$)
or diagonal ($s_d$) shift between the centers of the current loops
and the pinning centers, different vortex configurations, shown in
Fig.~2 with different symbols, appear to be stable after switching
off the magnetic field pulse. In the case of zero shifts $s_x$ and
$s_d$, no VAV lattice is formed in the film for current magnitudes
lower than $I_{\rm max}\approx 3.6$, even though at $I_{\rm max}<
3.6$ the applied magnetic field pulse can lead to strong local
suppressions of the order parameter. Interestingly, a moderate
nonzero shift $s_x$ facilitates nucleation and spatial separation of
vortices and antivortices from these suppressions. As follows from
Fig.~2, this can result in the formation of a stable VAV lattice
even for current magnitudes $I_{\rm max}$ slightly below 3.6. A
similar, but somewhat less pronounced effect, is seen to be caused
by a diagonal shift $s_d$.
\begin{figure}
\centering
\includegraphics*[width=0.95\linewidth]{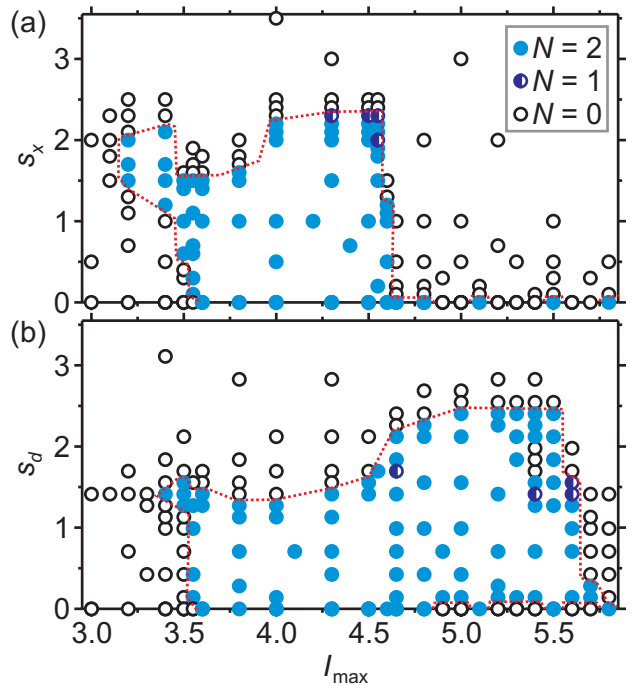}
\caption{Results for stabilization of a VAV lattice in the structure
shown in Fig.~1(a) for different current magnitudes $I_{\rm max}$
and shifts $s_x$ (a) or $s_d$ (b) between the centers of the current
loops and the pinning centers. The dotted lines approximately
indicate the boundaries between the ranges of parameters which
correspond to stabilization of a VAV lattice with two ($N = 2$,
light blue filled circles) or one ($N = 1$, dark blue semi-filled
circles) VAV pairs per unit cell, and the ranges of parameters which
correspond to eventual annihilation of pairs ($N = 0$, empty
circles). \label{Figure2}}
\end{figure}

At $3.6 < I_{\rm max} <4.6$, an antiferromagnetic lattice is
stabilized both at zero shifts $s_x$ or $s_d$ and also at moderate
nonzero values of these shifts. Our calculations show that for this
range of $I_{\rm max}$ the vortex pattern created at the end of the
pulse plateau ($t= t_2$) has exactly one single flux quantum vortex
or antivortex near the center of each current loop, thus forming a
VAV checkerboard lattice. Taking as a unit cell the square with side
$L$ [see Fig.~1(a)], this leads to two VAV pairs per unit cell. At
moderate shifts $s_x$ or $s_d$, when switching off the applied
magnetic field, this lattice simply shifts to the pinning-site
array. However, when the distance between the centers of the loops
and the pinning sites exceeds a critical value, complete VAV
annihilation, rather than the antiferromagnetic-lattice
stabilization, occurs after switching off the applied magnetic
field.

At larger $I_{\rm max}$, i.e. for higher magnetic fields, each
current loop generates more than one VAV pairs due to the
application of the magnetic field pulse. As a result, a giant vortex
(antivortex), which corresponds to two or more flux
quanta~\cite{Moshchalkov96,Moshchalkov98}, or a dense cluster of
single flux quantum vortices (antivortices) is accommodated by most
current loops at the end of the pulse plateau. Ramping down the
applied field during the time interval $(t_2,\ t_3)$ leads to
dissociation of these giant vortices (antivortices) or vortex
(antivortex) clusters into single-$\Phi_0$ vortices (antivortices).
The further evolution of the system is determined by an intricate
interplay of the vortex-(anti)vortex interactions as well as the
interaction of vortices and antivortices with the pinning potentials
and the gradually decreasing applied inhomogeneous magnetic field.
This leads to a rather complicated recombination dynamics of the
vortices and antivortices. After that, generally one of two outcomes
is realized: either we again obtain a stable VAV checkerboard
pattern with two VAV pairs per unit cell as before, or full
recombination has taken place. Remarkably, a moderate diagonal shift
$s_d$ of the current loops with respect to the pinning sites can
promote the antiferromagnetic-vortex-lattice stabilization for this
range of $I_{\rm max}$. No such effect is found for a shift $s_x$
along the side of the unit cell. The following qualitative
difference between diagonal and horizontal shifts seems to be
important: in the former case, the pinning sites lie on the lines
that connect loops with the same direction of current, while in the
latter case the pinning cites are shifted from a current loop
towards a loop with the opposite current direction. As further seen
from Fig.~2, besides the two ``limiting cases'' (stabilization of a
lattice with two VAV pairs per unit cell or full annihilation of
pairs), sometimes an ``intermediate'' configuration is observed for
parameter values near the boundary between the regions corresponding
to the aforementioned two ``limits''. This configuration has one VAV
pair per unit cell, and an example of this pattern will be given
below.

Large shifts between the pinning lattice and the current loop
lattice impede the stabilization of the VAV lattice. In order to
avoid this, the density of artificial pinning sites can be increased
so that a pinning site is always nearby for the vortices generated
by a current loop. An example of such a structure is shown in
Fig.~3(a). As in the cases analyzed above, stable antiferromagnetic
vortex lattices in this structure may contain either two VAV pairs
per unit cell or one VAV pair per unit cell. The latter is
illustrated by Fig.~3b, showing the evolution of the magnetic field
patterns, corresponding to vortices and antivortices, for the
particular set of parameters, $\delta = 0.09$, $d_1=1$, $D=2$,
$I_{\rm max}=5.5$, $t_1=10$, $t_2=160$, and $t_3=260$. By the end of
the magnetic-pulse plateau ($t = 160$), giant vortices
(antivortices), carrying two flux quanta, are formed at each current
loop. When decreasing the applied magnetic field, these giant
vortices and antivortices still persist up to $t \approx 244$ [see
Fig.~3(b)]. However, with further reduction of the applied field (at
$t \approx 252$), they dissociate into single-$\Phi_0$ vortices and
antivortices, which then partially recombine [see the snapshots for
$t = 254$ and 255 in Fig.~3(b)]. Since the pinning potentials in the
structures under consideration are relatively weak, they do not
prevent the corresponding motion of vortices and antivortices. This
means that these pinning centers can provide fine tuning in
stabilizing a state with coexisting vortices and antivortices. The
crucial prerequisite for such a stabilization is mutual cancelation
of vortex-vortex and vortex-antivortex interactions in a regular
antiferromagnetic VAV lattice. In the case under consideration, the
order parameter finally settles into a stable state with one VAV
pair per unit cell, corresponding to periodic VAV chains, separated
from each other by a distance $L$ [see the corresponding magnetic
field distribution at $t = 350$ in Fig.~3(b)]).
\begin{figure}
\centering
\includegraphics*[width=0.95\linewidth]{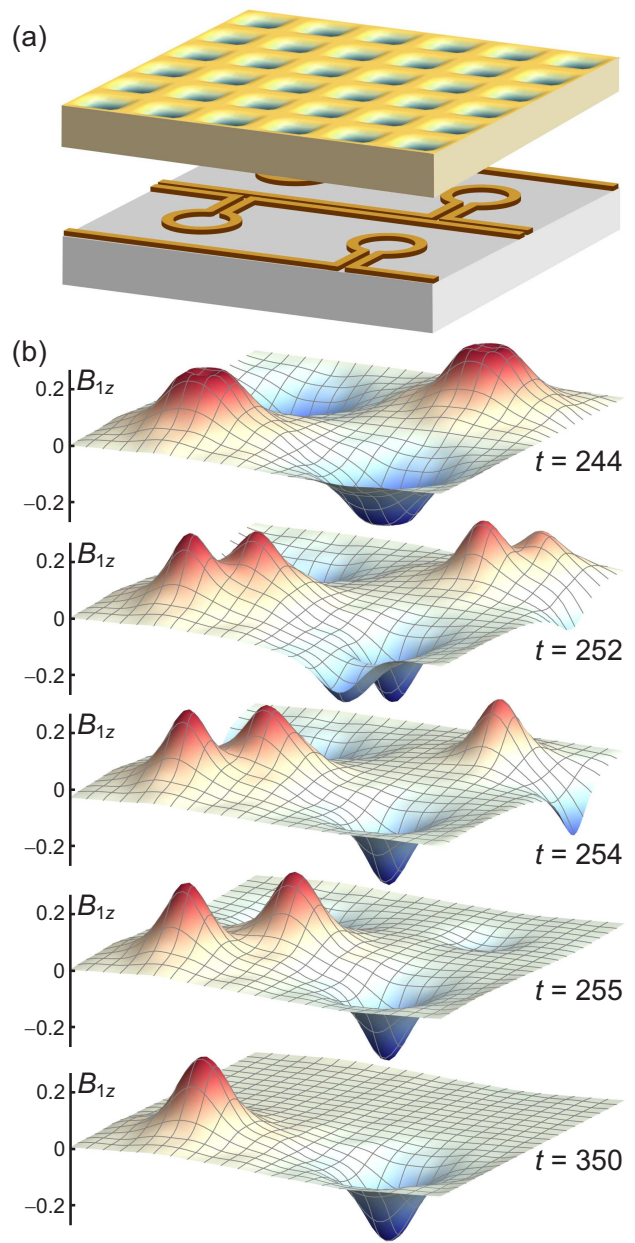}
\caption{(a) Unit cell with four current loops and a relatively high
density of pinning centers in the superconductor film. (b) Snapshots
of the magnetic field $B_{1z}$ induced by the superconducting film
at different time moments $t$. \label{Figure3}}
\end{figure}

Each current loop, independently of the sense of the current in it,
can generate both vortices and antivortices in the superconductor
film. This suggests that a stable VAV lattice can be formed also in
a periodic structure where the current direction is the same for all
the current loops. The results of our calculations for those
structures (see supporting information) confirm such a possibility.
Moreover, these results demonstrate that (at least for certain sets
of the relevant parameters) stabilization of a VAV lattice can be
possible even when the number of vortices and antivortices, induced
by each current loop in the course of the magnetic field pulse, is
much larger than one. At the same time, we find that the most
promising and reliable regime for the experimental realization of
the imprinted antiferromagnetic vortexantivortex lattices that
remain stable after the external magnetic field is switched off,
corresponds to the generation of just one VAV pair by each current
loop.

Since the unit of length, used in our calculations, is proportional
to the coherence length $\xi$, the examples given above can
correspond to different geometric parameters, depending on a
particular superconducting material and temperature. Thus, for the
coherence length $\xi\approx 250$~nm found in Ref.~\cite{Silhanek11}
for an Al film at $T=0.63T_c$, the structure shown in
Fig.~\ref{Figure3} corresponds to the unit-cell size about
5.3~$\mu$m, the superconducting-layer thickness 100~nm, the diameter
of current loops 700 nm and the spacing between the loops and the
superconducting layer about 350 nm. For the depth and lateral size
of the indentations, which serve as pinning sites, we then obtain
the values about 30~nm and 600~nm, respectively. Of course, all the
indicated values would appear larger (smaller) for higher (lower)
temperatures.

Experimentally, stable imprinted VAV lattices, theoretically
predicted in this paper, can be directly visualized in nanopatterned
systems (similar to the one shown in Fig.~\ref{Figure1}) by using
scanning SQUID~\cite{kirtley03,tafuri04,carillo03,kirtley04} or Hall
probe~\cite{gutierrez12} microscopies.

To conclude, we have proposed the application of a pulse of
spatially periodic up and down magnetic fields to imprint a
vortex-antivortex lattice on a superconducting film. The
stabilization of the imprinted VAV pattern can be provided by an
artificial array of pinning centers, so that the antiferromagnetic
vortex pattern remains stable in zero magnetic field, after the
magnetic field pulse is turned off. The presence of pinning centers
(especially, of their periodic arrays) in the superconducting layer
is a necessary but -- in general -- not sufficient condition for the
antiferromagnetic-vortex-lattice stabilization: it is sensitive also
to the magnitude of the applied magnetic field pulse as well as to
its spatial and temporal shape. Nevertheless, our results clearly
demonstrate that the relevant parameters, which assure creation of
stable antiferromagnetic vortex lattices with a relatively high
density of vortices and antivortices, can vary in a rather broad
range. Moreover, the stabilization of the antiferromagnetic vortex
lattices is even easier in presence of a denser periodic array of
artificial pinning centers.

\begin{acknowledgments}
This work was supported by Methusalem funding by the Flemish
government, the Flemish Science Foundation (FWO-Vl), in particular
FWO projects G.0356.05, G.0115.06, G.0370.09N, and G.0115.12N, the
Scientific Research Community project WO.033.09N, the Belgian
Science Policy, and the ESF NES network.
\end{acknowledgments}

\end{document}